\documentclass[12pt]{article}      
\usepackage{epsfig,multirow,setspace}

\AtEndOfClass{\RequirePackage{times}}
\usepackage{mathtools}
\usepackage{amsthm,amsmath,amssymb,colonequals,multirow,graphicx,epstopdf}
\usepackage{caption}
\usepackage{color}
\usepackage{mwe}

\usepackage[caption=false,font=footnotesize,labelfont=sf,textfont=sf]{subfig}
\usepackage{fancyhdr}
\usepackage{amsfonts, amsmath, amssymb, marvosym, upgreek, mathrsfs, dsfont} 
\usepackage{array,graphicx}
\usepackage{booktabs}
\usepackage{pifont}
\usepackage{setspace}
\usepackage{colonequals} 
\usepackage[authoryear]{natbib}
\usepackage{longtable}

\renewcommand\harvardyearright[1]{.}

\newcommand{\vecX}{\mathbf{X}}

\newcommand{\vecU}{\mathbf{U}}
\newcommand{\vecI}{\mathbf{I}}

\newcommand{\veczero}{\mathbf{0}}
\newcommand{\vecmu}{\mbox{\boldmath$\mu$}}

\newcommand{\vecLambda}{\mbox{\boldmath$\Lambda$}}

\newcommand{\matepsilon}{\mbox{\boldmath$\epsilon$}}
\newcommand{\matPsi}{\mbox{\boldmath$\Psi$}}

\newcommand{\bI}{\mathbf{I}}
\newcommand{\bP}{\mathbf{P}}
\newcommand{\bQ}{\mathbf{Q}}
\newcommand{\bS}{\mathbf{S}}
\newcommand{\bU}{\mathbf{U}}
\newcommand{\bW}{\mathbf{W}}
\newcommand{\bX}{\mathbf{X}}
\newcommand{\bY}{\mathbf{Y}}
\newcommand{\bZ}{\mathbf{Z}}

\newcommand{\bm}{\mathbf{m}}
\newcommand{\bmm}{\mathbf{m}}
\newcommand{\bu}{\mathbf{u}}

\newcommand{\bx}{\mathbf{x}}
\newcommand{\by}{\mathbf{y}}

\newcommand{\bpi}{\boldsymbol{\pi}}

\newcommand{\bbeta}{\boldsymbol{\beta}}
\newcommand{\bTheta}{\boldsymbol{\Theta}}
\newcommand{\bmu}{\boldsymbol{\mu}}
\newcommand{\bSigma}{\boldsymbol{\Sigma}}
\newcommand{\bLambda}{\boldsymbol{\Lambda}}
\newcommand{\bPsi}{\boldsymbol{\Psi}}

\newcommand{\tr}{\,\mbox{tr}}
\newcommand{\diag}{\,\mbox{diag}}

\DeclareMathOperator{\Ev}{\mathbb{E}}

\newcommand{\real}{\text{\rm I\hspace{-0.6mm}R}}

\begin{document}

\title{Finite Mixtures of Multivariate Poisson-Log Normal Factor Analyzers for Clustering Count Data}

\author{Andrea Payne\footnote{Co-first authors}~\footnote{School of Mathematics and Statistics, Carleton University, Ottawa, ON, Canada.}\qquad\qquad 
Anjali Silva\footnotemark[1]~\footnote{Princess Margaret Cancer Centre - University Health Network, Toronto, ON, Canada; University of Toronto Libraries, Toronto, ON, Canada. }\qquad\qquad Steven~J.~Rothstein\footnote{Department of Molecular and Cellular Biology, University of Guelph, Guelph, ON, Canada.}\\\hspace{-0.4in}Paul~D.~McNicholas\footnote{Department of Mathematics and Statistics, McMaster University, Hamilton, ON, Canada.} \qquad\qquad Sanjeena Subedi\footnotemark[2]}
\date{}

\maketitle

\begin{abstract}
A mixture of multivariate Poisson-log normal factor analyzers is introduced by imposing constraints on the covariance matrix, which resulted in flexible models for clustering purposes. In particular, a class of eight parsimonious mixture models based on the mixtures of factor analyzers model are introduced. Variational Gaussian approximation is used for parameter estimation, and information criteria are used for model selection. The proposed models are explored in the context of clustering discrete data arising from RNA sequencing studies. Using real and simulated data, the models are shown to give favourable clustering performance. The GitHub R package for this work is available at \texttt{https://github.com/anjalisilva/mixMPLNFA} and is released under the open-source MIT license.
\end{abstract}

\section{Introduction}\label{sec:intro}
Model-based clustering is a technique that utilizes finite mixture models to cluster data  \citep{wolfe1965, mclachlanandbasford1998, mclachlan2000, mcnicholas2016}. The general distribution function for mixture models can be given as 
\begin{align*}
    f( \mathbf{y}| \pi_1, \ldots, \pi_G,\boldsymbol{\vartheta}_1, \ldots, \boldsymbol{\vartheta}_G )= \sum_{g=1}^{G} \pi_g f_g( \mathbf{y} | \boldsymbol{\vartheta}_g),
\end{align*}
\noindent where $G$ is the total number of clusters, $f_g(\cdot)$ is the distribution function with parameters $\boldsymbol{\vartheta}_g$, and $ \pi_g>0$ is the mixing weight of the $g^{\text{th}}$ component such that $\sum_{g=1}^G\pi_g=1$. Mixture model-based clustering methods can be over-parameterized in high-dimensional spaces, especially as the number of clusters increases. Subspace clustering allows for clustering data in low-dimensional subspaces while utilizing information from all the dimensions and by introducing restrictions to mixture parameters \citep{bouveyron2012}. Restrictions are introduced to the model parameters to obtain parsimonious models, which provide various possible covariance structures while reducing the number of parameters to fit.

Clustering multivariate count data is of interest in a variety of fields, including, but not limited to, bioinformatics, text mining, and sports analytics \citep{subedi2020parsimonious}. Specifically within bioinformatics, this form of clustering has become increasingly popular as researchers aim to sort data into homogeneous subgroups so differing characteristics can be identified and analyzed \citep{subedi2020parsimonious}. Many novel clustering methods have been proposed in recent years; \cite{de2008} compare these methods to more traditional clustering methods, finding that a mixture model-based clustering approach outperformed the rest in the context of clustering cancer tissues for subtype identification. In the years since this publication, further research has proposed advancements to this approach and new methods for mixture model clustering for variety of data types \citep{bouveyron2012, gollini2014, subedi2014, vrbik14, subedi2015, browne2015, dang2015, kosmidis2016, tortora19, subedi2020, subedi2020parsimonious}. These proposed methods focus on a variety of distributions and methods of parameter estimation. 

RNA sequencing (RNA-seq) is a specific area of interest within the field of bioinformatics for applications of model-based clustering. RNA-seq is used to determine the transcriptional dynamics of a biological system by measuring the expression levels of many genes \citep{wang2009, roberts2011}. This technique provides counts of reads that can be mapped back to a biological entity. \cite{di2011} note that negative binomial-based models can account for variability in RNA-seq data while also requiring less statistical power. As such, negative binomial models have been utilized to cluster RNA-seq data as well \citep{si2013}. However, as stated by \cite{silva2019}, clustering methods based on a univariate distribution such as negative binomial distribution can struggle with clustering multivariate RNA-seq data as such models assume independence between variables. Hence, in this paper, we will develop a method based on multivariate Poisson-log normal distributions \citep{aitchison1989,silva2019} which can account for overdispersion and also accommodate positive and negative correlations.

The factor analysis model by \citet{spearman1904} assumes that a $d$-dimensional vector of observed variables, $\mathbf{\vecX}$, can be modelled by a $K$-dimensional vector of latent factors, where $K < d$. As a result, the factor analysis model is useful in modelling the covariance structure of high-dimensional data using a small number of latent variables. The mixture of factor analyzers model was later introduced by \citet{ghahramani1997} and this model can perform clustering and local dimensionality reduction within each cluster concurrently. Consider $n$ independent $d$-dimensional continuous variables $\vecX_1, \ldots, \vecX_n$, which come from a heterogeneous population with $G$ subgroups. In the mixture of factor analyzers framework, $\vecX_i$ is modelled as
\begin{equation*}
\vecX_i = \vecmu_g + \vecLambda_g\vecU_{ig}+ \matepsilon_{ig} \text{ with probability } \pi_{g},
\end{equation*} 
for $i=1,\ldots,n$ and $g=1,\ldots,G$. Here, $\vecmu_g$ is a $d \times 1$ vector of $g$th component mean, $\vecLambda_g$ is a $d \times K$ matrix of $g$th component factor loadings, $\vecU_{ig} \sim \mathcal{N}_K(\veczero,\vecI_K)$ is a $K \times 1$ vector of $g$th component latent factors, and $\matepsilon_{ig}\sim\mathcal{N}_d(\veczero,\matPsi_g)$ is a $d \times 1$ vector of $g$th component errors with $\matPsi_g = \text{diag}(\psi_{g1},\ldots,\psi_{gd})$. Note that the $\vecU_{ig}$ are independently distributed and are independent of the $\matepsilon_{ig}$, which are also independently distributed. Under this  model, the density of $\vecX_i$ from a mixture of factor analyzer models is 
\begin{equation*}
f(\mathbf{x}_i \mid \boldsymbol{\Theta}) = \sum_{g=1}^G \pi_g f_g(\mathbf{x}_i \mid \vecmu_g, \vecLambda_g \vecLambda_g^{\prime} + \matPsi_g),
\end{equation*} 
where, $\boldsymbol{\Theta} = (\pi_1, \ldots, \pi_G, \vecmu_1, \ldots, \vecmu_G, \mathbf{\Lambda}_1, \ldots, \mathbf{\Lambda}_G, \mathbf{\Psi}_1, \ldots, \mathbf{\Psi}_G)$. Further, $\vecX_i| \mathbf{u}_{ig} \sim \mathcal{N}_d(\vecmu_g + \vecLambda_g \mathbf{u}_{ig}, \matPsi_g)$. It should be noted that the $\mathbf{\Lambda}_g$ is not uniquely defined for $K > 1$. Therefore the $K$-dimensional space in which the factors lie can be determined, but the directions of these factors cannot be determined.  However, this does not affect the clustering algorithm, because $\mathbf{\Lambda}_g \mathbf{\Lambda}_g^{\prime}$ is unique. The number of free $g^{th}$ component covariance parameters that are reduced using the factor analysis model is
\begin{equation*}
\frac{1}{2} d(d+1) - \Big[ dK + d - \frac{1}{2}K(K-1) \Big] = \frac{1}{2} \Big[ (d-K)^2 - (d+K) \Big],
\end{equation*} 
given that $(d-K)^2 > (d+K)$ \citep{lawley1962, mcnicholas2016}.

In 2008, this work was extended and a family of eight parsimonious Gaussian mixture models \citep[PGMMs;][]{mcnicholas2008} were introduced with parsimonious covariance structures. The PGMM family arises by considering the general mixture of factor analyzers model ($\mathbf{\Sigma}_g = \mathbf{\Lambda}_g  \mathbf{\Lambda}_g^{\prime} + \mathbf{\Psi}_g $) and by allowing the constraints $\mathbf{\Lambda}_g = \mathbf{\Lambda}$, $\mathbf{\Psi}_g = \mathbf{\Psi}$, and the isotropic constraint $\mathbf{\Psi}_g = \psi_{g}\vecI_d$. These covariance structures can have as few as $dK - K(K - 1)/2 + 1$ free parameters or as many as $G[dK - K(K - 1)/2 + d]$ free parameters. With the introduction of the factor analysis structure, the number of covariance parameters is linear in data dimensionality, thus making this family well suited for analysis of high-dimensional data \citep{mcnicholas2010}. The constraints allow for assuming a common structure in the component covariance matrix $\mathbf{\Sigma}_g$ between clusters, if appropriate, and this enables a parsimonious model. 

Previously, a model-based clustering methodology using a mixture of multivariate Poisson-log normal distributions \citep[MPLN;][]{aitchison1989} was developed to analyze multivariate count measurements from RNA-seq studies \citep{silva2019}. A $d$-dimensional random variable following a $G$-component mixtures of MPLN distribution is said to have a total of $G-1$ + $Gd$ + $Gd(d+1)/2$ free parameters. Here, $G-1$ parameters are contributed by the mixing proportions, $Gd$ from the means and $Gd(d+1)/2$ from the covariance matrices. Since the largest contribution is through the covariance matrices, it is a natural focus for the introduction of parsimony. 

In this work, a family of mixtures of MPLN factor analyzers that is analogous to the PGMM family is developed, by considering the constraints $\mathbf{\Lambda}_g = \mathbf{\Lambda}$ and $\mathbf{\Psi}_g = \mathbf{\Psi}$. This family is referred to as the parsimonious mixtures of the MPLN factor analyzers family (MPLNFA). The proposed model simultaneously performs factor analysis and cluster analysis, by assuming that the discrete observed data have been generated by a factor analyzer model in the continuous latent variables. Details of parameter estimation are provided, and both real and simulated data illustrations are used to demonstrate the clustering ability.

\section{Methodology}
 \subsection{Mixtures of Factor Analyzer}
Suppose $\mathbf{Y}_{ij}$ denotes the observed counts of the $i^{th}$ sample for the $j^{th}$ variable and $\mathbf{X}_{ij}$ denotes the associated underlying latent variable. To develop a mixture of factor analyzers, we assume the following hierarchical structure with the $K$-dimensional latent factor $\mathbf{U}_i$:
\begin{align*}
Y_{ij}&\mid X_{ij}=x_{ij}\sim \text{Poisson}\left(\exp\{x_{ij}+\log{C_i}\}\right),\\
\bX_i&\mid \bU_i=\bu_i \sim N(\bmu_g+\bLambda_g \bu_i,\bPsi_g),\\
\bU_i &\sim N(\mathbf{0}_K,\bI_K).
\end{align*}
Here, $C_i$ is a fixed known constant representing the normalized library sizes added to account for the differences in the library sizes of the $i^{th}$ sample.

In model-based clustering, we also have unobserved component membership indicator variable $\mathbf{Z}$ such that $Z_{ig}=1$ if the observation $i^{th}$ belongs to group $g$ and $Z_{ig}=0$ otherwise. 
Hence, the complete data now comprises observed expression levels $\by$, underlying latent variable $\bX$, and unknown group membership $\mathbf{Z}$. The complete-data likelihood can therefore be written as:
\begin{align*}
L(\boldsymbol{\vartheta}) &= \prod_{g=1}^G\prod_{i=1}^n\left[\pi_g \left\{\prod_{j=1}^d f(y_{ij}\mid x_{ij},C_i)\right\}~f(\bx_i\mid \bu_i,\bmu_g,\bLambda_g,\bPsi_g)~ f(\bu_i)\right]^{z_{ig}},
\end{align*}
where $\boldsymbol{\vartheta}$ denotes all the model parameters. The complete data log-likelihood can therefore be written as:
\begin{align*}
l_c&(\boldsymbol{\vartheta})= \\
&\sum_{g=1}^G\sum_{i=1}^n z_{ig} \left[\log \pi_g + \left\{\sum_{j=1}^d \log f(y_{ij}\mid x_{ij},k_j)\right\}+ \log f(\bx_i\mid \bu_i,\bmu_g,\bLambda_g,\bPsi_g) + \log f(\bu_i)\right],
\end{align*}
To utilize the expectation-maximization (EM) framework for parameter estimation, we require 
$\Ev(Z_{ig}\bX_i\mid \by_i,\bu_i, \bmu_g,\bLambda_g,\bPsi_g)$, $\Ev(Z_{ig}\bX_i\bX_i'\mid\by_i,\bu_i, \bmu_g,\bLambda_g,\bPsi_g)$, $\Ev(Z_{ig}\bU_i\mid \by_i,\bx_i, \bmu_g,\bLambda_g,\bPsi_g)$ and $\Ev(Z_{ig}\bU_i\bU_i'\mid\by_i,\bx_i, \bmu_g,\bLambda_g,\bPsi_g)$. To compute these quantities, we need the posterior distribution of $\bX_i\mid\by_i,\bu_i$ which does not have a closed form. Thus, parameter estimation relies on the Markov chain Monte Carlo expectation-maximization (MCMC-EM) approach which can be computationally intensive.

\subsection{Variational approximation of mixtures of MPLN factor analyzers}\label{sec:method3}

\cite{subedi2020parsimonious} proposed a framework for parameter estimation utilizing variational Gaussian approximation (VGA) for MPLN-based mixture models. Variational approximation \citep{wainwright2008} is an approximate inference technique that uses a computationally convenient approximating density in place of a more complex but `true' posterior density obtained by minimizing the Kullback-Leibler (KL) divergence between the true and the approximating densities. Markov chain Monte Carlo expectation-maximization (MCMC-EM) has also been used for parameter estimation of MPLN-based mixture models, but VGA was shown to be computationally efficient \citep{silva2023}. Here, we propose a mixture of Poisson log-normal distributions with factor analyzers that utilize a variational EM framework for parameter estimation.

Parameter estimation for a mixture of factor analyzers is typically done using an alternating expectation conditional maximization (AECM) framework that assumes different specifications of missing data at different stages. Here, we will adopt a similar notion resulting in a two-stage iterative EM-type approach.

\subsection{Stage 1}
In Stage 1, we treat $\bX$ and $\bZ$ as missing such that
\begin{align*}
Y_{ij}&\mid X_{ij}=x_{ij}\sim \text{Poisson}\left(\exp\{x_{ij}+\log{C_i}\}\right)\\
\bX_i&\sim N(\bmu_g,\bSigma_g),
\end{align*}
where $\bSigma_g=\bLambda_g\bLambda_g^T+\bPsi_g$.Therefore, the component-specific marginal density of the observed data $\by_i$ can be written as
\vspace{-0.05in}
\begin{align*}
 f(\by_i\mid Z_{ig}=1,\bmu_g,\bSigma_g)=\int_{\real^d} \left[ \prod _{j=1}^d f_p(y_{ij}\mid x_{ij},C_i, Z_{ig}=1)\right] ~f_N(\bx_i\mid Z_{ig}=1,\bmu_g,\bSigma_g)~d\bx_i,
\end{align*}
where $x_{ij}$ and $y_{ij}$ are the $j^{th}$ element of $\bx_i$ and $\by_i$ respectively, $f_p(\cdot)$ is the probability mass function of the Poisson distribution with mean $e^{x_{ij}+\log C_j}$ and $f_N(\cdot)$ is the probability density function of  $d$-dimensional Gaussian distribution with mean $\bmu_g$ and covariance $\bSigma_g$. Note that the marginal distribution of $\bY$ involves multiple integrals and cannot be further simplified. The complete-data log-likelihood using the marginal density of $\by_i\mid Z_{ig}=1$ can be written as:
$$l(\boldsymbol{\vartheta})=\sum_{g=1}^G\sum_{i=1}^n z_{ig}\log \pi_g+ \sum_{g=1}^G\sum_{i=1}^n z_{ig}\log f(\by_{i}\mid Z_{ig}=1, \bmu_g,\bSigma_g).$$
As the marginal of $\bY_i\mid Z_{ig}=1$ cannot be simplified, we use variational Gaussian approximation to approximate the marginal of $\bY_i$ from component $g$ in the mixtures of MPLN distributions. Suppose, we have an approximating density $q(\bx_{ig})$, we can write
\begin{align*}
\log f(\by_i\mid Z_{ig}=1) &= F(q_{ig},\by_i) + D_{KL}(q_{ig}\|f_{ig}),
\end{align*}

\noindent where $D_{KL}(q_{ig}\|f_{ig})= \int_{\real^d} q(\bx_{ig}) \log \frac{q(\bx_{ig})}{f(\bx_{i}\mid\by_i,Z_{ig}=1)} d\bx_{ig}$ is the Kullback-Leibler (KL) divergence between $f(\bx_{i}\mid \by_i,Z_{ig}=1)$ and approximating distribution $q(\bx_{ig})$, and 
$$F(q_{ig},\by_i)=\int_{\real^d} \left[\log f(\by_i,\bx_i\mid Z_{ig}=1) - \log q(\bx_{ig})\right] q(\bx_{ig})  d\bx_{ig},$$ is our evidence lower bound (ELBO) for each observation $\by_i$. Details can be found in Appendix \ref{app:vae}.
\noindent The complete data log-likelihood of the mixtures of MPLN distributions can be written as:
\begin{align*}
l_c(\boldsymbol{\vartheta}\mid \by) &= \sum_{g=1}^G \sum_{i=1}^n z_{ig} \log \pi_g +  \sum_{g=1}^G \sum_{i=1}^n z_{ig} \left[ F(q_{ig},\by_i) + D_{KL}(q_{ig}\|f_{ig}) \right].
\end{align*}

\noindent In VGA, $q(\bx_{ig})$ is assumed to be a Gaussian distribution. Assuming $q(\bx_{ig}) = \mathscr{N}(\mathbf{m}_{ig}, \mathbf{S}_{ig})$, the ELBO for each observation $\by_i$ becomes
\begin{align*}
F(q_{ig},\by_i) &= \\
&-\frac{1}{2}(\bmm_{ig}-\bmu_g)'\bSigma_g^{-1}(\bmm_{ig}-\bmu_g)-\frac{1}{2} \tr(\bSigma_g^{-1}\mathbf{S}_{ig})+\frac{1}{2} \log |\mathbf{S}_{ig}| -\frac{1}{2} \log |\bSigma_g| + \frac{d}{2} \\
&+ \bmm_{ig}'\by_i +\sum_{j=1}^d\left(\log C_i\right)y_{ij}-   \sum_{j=1}^d \left\{e^{\log C_i+m_{igj}+\frac{1}{2}S_{ig,jj}}+\log(y_{ij}!)\right\}.
\end{align*}
Thus, to minimize the KL divergence between the true and approximating density, we find the values of the variational parameters $\bm_{ig}$ and $\bS_{ig}$ that maximize our ELBO. Similar to \cite{arridge2018} and \cite{subedi2020parsimonious}, estimation of the mean ($\bmm_{ig}$) and variance ($\mathbf{S}_{ig}$) can be obtained via Newton's method and fixed-point method.

Therefore, in Stage 1, we update $\Ev(Z_{ig}\mid \by_i)$, variational parameters $\bmm_{ig}$  and $\mathbf{S}_{ig}$, and model parameters $\pi_g$ and $\bmu_g$.
\begin{enumerate}
\item Conditional on the variational parameters $\bmm_{ig},\mathbf{S}_{ig}$ and on $\bmu_g$ and $\bSigma_g$, the  $\Ev(Z_{ig})$ is computed. Given $\bmu_g$ and $\bSigma_g$, 
\begin{align*}
\Ev(Z_{ig}\mid \by_i)=\frac{\pi_g f(\by\mid \bmu_g,\bSigma_g)}{\sum_{h=1}^G\pi_h f(\by\mid \bmu_h,\bSigma_h)}.
\end{align*}
Note that this involves the marginal distribution of $\bY$ which is difficult to compute. Hence, similar to \citet{subedi2020parsimonious}, we use an approximation of $\Ev(Z_{ig})$ where we replace the marginal density of the exponent of ELBO such that 
\begin{align*}
\widehat{Z}_{ig}\stackrel{\text{def}}{=}\frac{\pi_g \exp\left[F\left(q_{ig},\by_i\right) \right]}{\sum_{h=1}^G\pi_h \exp\left[F\left(q_{ih},\by_i\right) \right]}.
\end{align*}

\item Update the variational parameters $\bmm_{ig}$ and $\bS_{ig}$ as following:
\begin{enumerate}
 \item Fixed-point method for updating $\mathbf{S}_{ig}$ is
\begin{align*}
\hspace{-0.65in} \mathbf{S}_{ig}^{(t+1)}=\left\{\bSigma_g^{-1}+ \bI \odot \mathbf{exp}\left[\log C_i+\bmm_{ig}^{(t)}+\frac{1}{2} \diag \left(\mathbf{S}_{ig}^{(t)}\right)\right] \mathbf{1}_d'  \right\}^{-1}
\end{align*}
where the vector function $\mathbf{exp}\left[ \mathbf{a} \right] = (e^{a_1}, \ldots, e^{a_d})'$ is a vector of exponential each element of the $d$-dimensional vector $\mathbf{a}$, $\mbox{diag}( \mathbf{S} ) = ( \mathbf{S}_{11} \ldots,  \mathbf{S}_{dd})$ puts the diagonal elements of the $d\times d$ matrix  $\mathbf{S}$ into a d-dimensional vector, $\odot$ the Hadmard product and $\mathbf{1}_d$ is a d-dimensional vector of ones ;
\item Newton's method to update $\bmm_{ig}$ is
\begin{align*}
 \bmm_{ig}^{(t+1)}&=\bmm_{ig}^{(t)}\! -\! \mathbf{S}_{ig}^{(t+1)}\! \left\{ \! \by_i - \mathbf{exp}\left[\log C_i+\bmm_{ig}^{(t)}\! +\! \frac{1}{2}\diag\left(\mathbf{S}_{ig}^{(t+1)}\right)\right]\! \right.\\
 &\left. -\! \bSigma_g^{-1} \left(\bmm_{ig}^{(t)}-\bmu_g\right)\! \right\}\!.
\end{align*}
\end{enumerate}
\item Given $\widehat{Z}_{ig}$ and the variational parameters $\bmm_{ig}$ and $\mathbf{S}_{ig}$, the parameters $\bpi$ and $\bmu_g$ are updated as:
\begin{align*}
\widehat{\pi}_g&=\frac{\sum_{i=1}^n \widehat{Z}_{ig}}{n},\\
\widehat{\bmu}_g& = \frac{\sum_{i=1}^n \widehat{Z}_{ig} \bmm_{ig}^{(t+1)}}{\sum_{i=1}^n \widehat{Z}_{ig}}.
\end{align*}
\end{enumerate}

\subsection{Stage 2}
In Stage 2, we treat $\bX$, $\bU$ and $\bZ$ as missing such that
\begin{align*}
Y_{ij}&\mid X_{ij}=x_{ij}\sim \text{Poisson}\left(\exp\{x_{ij}+\log{C_i}\}\right)\\
\bX_i&\mid \bU_i=\bu_i\sim N(\bmu_g+\bLambda_g \bu_i,\bPsi_g),\\
\bU_i&\sim N(\mathbf{0}_K,\bI_K).
\end{align*}
Here, we use $d$ for the dimensionality of the observed data and $K$ as the dimensionality of the latent variable.
Suppose, we have an approximating density $q_{\bx,\bu}(\bx_{ig},\bu_{ig})$, the marginal of the log of the probability mass function of $\bY_i\mid Z_{ig}=1$ can be written as:
\begin{align*}
\log f(\by_i\mid Z_{ig}=1) &= F(q_{ig},\by_i) + D_{KL}(q_{ig}\|f_{ig}),
\end{align*}

\noindent where $D_{KL}(q_{ig}\|f_{ig})= \int_{\real^p} \int_{\real^d} q_{\bx,\bu}(\bx_{ig},\bu_{ig}) \log \frac{q_{\bx,\bu}(\bx_{ig},\bu_{ig})}{f(\bx_{i},\bu_{i}\mid\by_i,Z_{ig}=1)} ~d\bx_{ig}~d\bu_{ig}$ is the Kullback-Leibler (KL) divergence between $f(\bx_{i},\bu_i \mid \by_i,Z_{ig}=1)$ and approximating distribution $q(\bx_{ig},\bu_{ig})$, and 
$$F(q_{ig},\by_i)=\int_{\real^p} \int_{\real^d} \Big[\log f(\by_i,\bx_i,\bu_i\mid Z_{ig}=1) - \log q_{\bx,\bu}(\bx_{ig},\bu_{ig})\Big] q_{\bx,\bu}(\bx_{ig},\bu_{ig})  ~d\bx_{ig}~d\bu_{ig},$$ is our evidence lower bound (ELBO) for each observation $\by_i$.

\noindent If we assume $q_{\bx,\bu}(\bx_{ig},\bu_{ig})$ to be factorable such that $q_{\bx,\bu}(\bx_{ig},\bu_{ig})=q_{\bx}(\bx_{ig})q_{\bu}(\bu_{ig})$,
\begin{align*}
F&(q_{ig},\by_i)= \\
&\int_{\real^p} \Bigg[\left\{\int_{\real^d} \Big[ \log f(\by_i \mid \bx_i,\bu_i, Z_{ig}=1) +\log f(\bx_i\mid \bu_i,Z_{ig}=1) -\log q_\bx(\bx_i)\Big]~q_{\bx}(\bx_{ig})~d\bx_{ig} \right\}\\
&+ \log f(\bu_i\mid Z_{ig}=1)  - \log q_{\bu}(\bu_{ig}) \Bigg] q_{\bu}(\bu_{ig}) ~d\bu_{ig}.
\end{align*}
Using the approximating density $q_\bx(\bx_{ig})$ from Stage 1 (i.e. $q_\bx(\bx_{ig}) = N (\bmm_{ig},\mathbf{S}_{ig})$) and assuming $q_\bu(\bu_{ig}) = N (\bP_{ig},\mathbf{Q}_{ig})$, we get

\begin{align*}
F(q_{ig},\by_i)&=\bmm_{ig}'\by_i + \sum_{j=1}^d\left(\log C_i\right)y_{ij}-   \sum_{j=1}^d \left\{e^{\log C_i+m_{igj}+\frac{1}{2}S_{ig,jj}}+\log(y_{ij}!)\right\} +\frac{1}{2} \log |\bS_{ig}| +\frac{d}{2}\\
&-\frac{1}{2}(\bmm_{ig}-\bmu_{ig})^T\bPsi_g^{-1}(\bmm_{ig}-\bmu_{ig})+(\bmm_{ig}-\bmu_g)^T\bPsi_g^{-1}\bLambda_g\bP_{ig}-\frac{1}{2}\bP_{ig}^T\bLambda_g^T\bPsi_g^{-1}\bLambda_g\bP_{ig}\\
&-\frac{1}{2}\tr(\bLambda_g^T\bPsi_g^{-1}\bLambda_g\bQ_{ig})-\frac{1}{2}\tr\left(\bPsi_g^{-1}\mathbf{S}_{ig}\right)  -\frac{1}{2} \log |\bPsi_g|-\frac{1}{2}\bP_{ig}^T\bP_{ig}+\frac{1}{2}\log | \bQ_{ig}^{-1}|\\
&-\frac{1}{2}\tr( \bQ_{ig}) + \frac{K}{2}.
\end{align*}
The variational parameters $\bP_{ig}$ and $\bQ_{ig}$ that maximize this ELBO will also minimize the KL divergence. Therefore, in Stage 2, we update variational parameters $\bP_{ig}$ and $\bQ_{ig}$, and model parameters $\bLambda_g$ and $\bPsi_g$.
\begin{enumerate}
\item The update for the variational parameters $\bP_{ig}$ and $\bQ_{ig}$ are:
\begin{align*}
\widehat{\bQ}_{ig}^{(t+1)}&=(\bI_p+\hat{\bLambda}_g^T\hat{\bPsi}_g^{-1}\hat{\bLambda}_g)^{-1}\\
&=\bI_p-\hat{\bbeta}_g\hat{\bLambda}_g (\text{after a bit of simplification where }\hat{\bbeta}_g=\hat{\bLambda}_g^T(\hat{\bPsi}_g+\hat{\bLambda}_g\hat{\bLambda}_g^T)^{-1}).\\
\widehat{\bP}_{ig}^{(t+1)}&=\widehat{\bQ}_{ig}^{(t+1)}\hat{\bLambda}_g^T\hat{\bPsi}^{-1}_g(\bmm_{ig}^{(t+1)}-\hat{\bmu}_{g})\\
&=\hat{\bbeta}_g(\bmm_{ig}^{(t+1)}-\hat{\bmu}_{g}).
\end{align*}
\item The update for $\bPsi_g$ and $\bLambda_g$ are obtained by repeating the following steps until convergence:
\begin{align}
\widehat{\bTheta}_g&=\bI_p-\hat{\bbeta}_g\hat{\bLambda}_g+\hat{\bbeta}_g\bW_g\hat{\bbeta}_g^T, \nonumber\\
\hat{\bbeta}_g&=\hat{\bLambda}_g^T(\hat{\bLambda}_g\hat{\bLambda}_g^T+\hat{\bPsi}_g)^{-1}, \nonumber \\
\hat{\bLambda}_g&=\bW_g\hat{\bbeta}^T\widehat{\bTheta}_g^{-1},\\
\hat{\bPsi}_g&=\diag\left(\bW_g-\hat{\bLambda}_g\hat{\bbeta}_g\bW_g+\frac{\sum_{i=1}^n \widehat{Z}_{ig}\bS_{ig}}{\sum_{i=1}^n\widehat{Z}_{ig}}\right),
\end{align}
where $\bW_{g}=\frac{\sum_{i=1}^n\widehat{Z}_{ig}(\bmm_{ig}^{(t+1)}-\hat{\bmu}_g)(\bmm_{ig}^{(t+1)}-\hat{\bmu}_g)^T}{\sum_{i=1}^n\widehat{Z}_{ig}}$Note: convergence is assumed when the difference in Frobenius norm of $\hat{\bLambda}_g$ and $\hat{\bPsi}_g$ from successive iterations are both less than $10^{-6}$.
\end{enumerate}

\section{Family of models}
To obtain a family of parsimonious models, constraints can be imposed on $\bLambda_g$ and $\bPsi_g$ as described in Table \ref{tabfam} that results in a family of 8 models. 
\begin{table}[!h]
\centering\small
\caption{Family of models based on the constraints on $\bLambda_g$ and $\bPsi_g$.}\label{tabfam}
\resizebox{\columnwidth}{!}{
\begin{tabular}{llllc} 
\hline
Model ID & Loading Matrix $\bLambda_g$  & Error Variance $\bPsi_g$ & Isotropic $\psi_g \mathbf{I}$ & Covariance parameters \\
\hline
UUU & unconstrained   & unconstrained & unconstrained & $ G\left[dK - K\left(K -1\right)/2\right] + Gd$ \\
UUC &  unconstrained   & unconstrained & constrained & $G\left[dK -K\left(K-1\right)/2\right] +G$\\
UCU &  unconstrained   & constrained & unconstrained & $G\left[dK -K\left(K -1\right)/2\right] + d$ \\
UCC & unconstrained   & constrained & constrained & $ G\left[dK -K\left(K-1\right)/2\right] + 1$ \\
CUU &  constrained   & unconstrained & unconstrained & $ \left[dK -K\left(K - 1\right)/2\right] + Gd$ \\
CUC &  constrained   & unconstrained & constrained & $ \left[dK -K\left(K- 1\right)/2\right] + G$ \\
CCU &  constrained   & constrained & unconstrained & $ \left[dK -K\left(K - 1\right)/2\right] + d$ \\
CCC &  constrained   & constrained & constrained & $ \left[dK -K\left(K - 1\right)/2\right] + 1$   \\
\hline
\end{tabular}
}
\end{table}

The estimation of $\bPsi_g$ and $\bLambda_g$ under various constraints are provided below:

\begin{itemize}
\item Under the isotropic constraint (i.e. $\bPsi_g = \psi_g \bI_d$), 
\begin{equation}
\hat{\psi}_g=\frac{1}{d}\tr\left(\bW_g-\hat{\bLambda}_g\hat{\bbeta}_g\bW_g+\frac{\sum_{i=1}^nz_{ig}\mathbf{S}_{ig}}{n_g}\right).
\end{equation}

\item Under the equal variance constraint (i.e., $\bPsi_g = \bPsi$),
\begin{align*}
\hat{\bPsi}&=\diag\left(\sum_{g=1}^G\frac{n_g}{n}\bW_g-\sum_{g=1}^G\frac{n_g}{n}\hat{\bLambda}_g\hat{\bbeta}_g\bW_g+\frac{1}{n}\sum_{g=1}^G\sum_{i=1}^nz_{ig}\mathbf{S}_{ig}\right).
\end{align*}

\item Under the equal Variance and isotropic constraint (i.e., $\bPsi_g=\psi\bI_d$),
\begin{align}
\hat{\psi}&=\frac{1}{d}\tr\left(\sum_{g=1}^G\frac{n_g}{n}\bW_g-\sum_{g=1}^G\frac{n_g}{n}\hat{\bLambda}_g\hat{\bbeta}_g\bW_g+\frac{1}{n}\sum_{g=1}^G\sum_{i=1}^nz_{ig}\mathbf{S}_{ig}\right).
\end{align}

\item Under the equal loading matrices constraint (i.e. $\bLambda_g = \bLambda$), the loading matrix cannot be solved directly and must be solved in a row-by-row manner as suggested by \cite{mcnicholas2008}. Hence,
\begin{align}
\hat{\lambda}_i&=\mathbf{r}_i\left(\sum_{g=1}^G\frac{n_g}{\hat{\Psi}_{g,(i)}} \hat{\bTheta}_g\right)^{-1},\label{common_lambda}
\end{align}
where $\lambda_i$ is the $i^{th}$ row of the matrix $\bLambda$, $\Psi_{g,(i)}$ is the $i^{th}$ diagonal element of $\bPsi_g$ and $\mathbf{r}_i$ is the $i^{th}$ row of the matrix $\sum_{g=1}^Gn_g\hat{\bPsi}_g^{-1}\bW_g\hat{\bbeta}_g^T$.

\end{itemize}

Full details of the parameter updates for all models can be found in Appendix \ref{app:para-update}.

%
%

\section{Results}
\subsection{Simulation Study}
Simulation studies were conducted to illustrate the ability to recover the true underlying parameters by the MPLNFA. Three simulation settings were utilized, each from a different model in the family of models. Within the three settings, three different numbers of clusters and latent factors were utilized and 100 datasets were generated with 1,000 observations per dataset under each setting.  The exact specifications of the simulation settings are summarized in Table ~\ref{settings_sim}.

\begin{table}[h!]
\centering
\caption{Summary of the simulation settings.}
\label{settings_sim}
\resizebox{0.8\textwidth}{!}{%
\begin{tabular}{ccccc}
\hline
Setting & Number of & Number of $n$ & Dimensions, $d$ & True Model \\
 & Simulated Datasets & Observations, $n$ & &  \\ \hline
1 & 100 & 1000 & 8 & UCC\\
2 & 100 & 1000 & 10 & CCC\\
3 & 100 & 1000 & 10 & UUU\\ \hline
\end{tabular}}\\
\resizebox{0.8\textwidth}{!}{%
\begin{tabular}{ccccc}
\hline
Setting & Number of  & Mixing Proportion, $\pi$ & Number of & True Model\\
 & Clusters, $G$ &  & Latent factors, $K$ & \\ \hline
1 & 4 & (0.11, 0.43, 0.24, 0.22) & 2 & UCC \\
2 & 2 & (0.32, 0.68) & 3 & CCC       \\
3 & 3 & (0.23, 0.44, 0.33) & 4 & UUU         \\ \hline
\end{tabular}}
\end{table}


\begin{table*}

\caption{Comparison of true and recovered parameters for all three simulation settings.}

\label{tab-parameter-recovery}
\resizebox{\textwidth}{!}{%
\begin{tabular}{c|c|c}
\hline
\multicolumn{3}{c}{Simulation Setting 1 - UCC} \\
\hline
\multirow{4}{*}{\textbf{Group 1}}  & $\boldsymbol{\mu}$ & (6, 3, 3, 6, 3, 6, 3, 3) \\
 & $\boldsymbol{\hat{\mu}}$ & (5.99, 3.00, 3.00, 6.00, 3.00, 6.00, 3.00, 3.00) \\   
 & Standard Deviation & (0.07, 0.10, 0.08, 0.13, 0.08, 0.07, 0.10, 0.08) \\
 & MSE($\boldsymbol{\Sigma}$) & 0.01 \\
\hline
\multirow{4}{*}{\textbf{Group 2}}  & $\boldsymbol{\mu}$ & (1, 3, 5, 1, 3, 5, 3, 5) \\
 & $\boldsymbol{\hat{\mu}}$ & (1.44, 2.85, 5.14, 1.42, 2.86, 5.15, 3.14, 4.86) \\   
 & Standard Deviation & (1.05, 0.36, 0.36, 1.04, 0.35, 0.35, 0.36, 0.35) \\
 & MSE($\boldsymbol{\Sigma}$) & 0.07 \\
\hline
\multirow{4}{*}{\textbf{Group 3}}  & $\boldsymbol{\mu}$ & (4, 2, 6, 4, 2, 6, 4, 4) \\
 & $\boldsymbol{\hat{\mu}}$ & (3.58, 2.16, 5.86, 3.59, 2.14, 5.86, 3.87, 4.13) \\   
 & Standard Deviation & (1.03, 0.34, 0.36, 1.05, 0.36, 0.36, 0.36, 0.35) \\
 & MSE($\boldsymbol{\Sigma}$) & 0.07 \\
\hline
\multirow{4}{*}{\textbf{Group 4}}  & $\boldsymbol{\mu}$ & (5, 3, 5, 3, 5, 3, 3, 5) \\
 & $\boldsymbol{\hat{\mu}}$ & (5.00, 2.99, 4.99, 3.00, 5.00, 3.01, 3.00, 5.00) \\   
 & Standard Deviation & (0.04, 0.05, 0.05, 0.05, 0.03, 0.04, 0.03, 0.06) \\
 & MSE($\boldsymbol{\Sigma}$) & 0.00 \\
\hline

\multicolumn{3}{c}{Simulation Setting 2 - CCC} \\
\hline
\multirow{4}{*}{\textbf{Group 1}}  & $\boldsymbol{\mu}$ & (6, 3, 3, 6, 3, 6, 3, 3, 6, 3) \\
 & $\boldsymbol{\hat{\mu}}$ & (6.00, 3.00, 3.00, 6.00, 3.00, 6.01, 3.00, 3.00, 6.00, 3.00) \\   
 & Standard Deviation & (0.05, 0.07, 0.05, 0.07 0.05, 0.05, 0.07, 0.06, 0.05, 0.06) \\
 & MSE($\boldsymbol{\Sigma}$) & 0.00 \\
\hline
\multirow{4}{*}{\textbf{Group 2}}  & $\boldsymbol{\mu}$ & (5, 3, 5, 3, 5, 5, 3, 5, 3, 5) \\
 & $\boldsymbol{\hat{\mu}}$ & (4.99, 3.01, 5.00, 3.00, 5.00, 5.00, 3.01, 5.00, 3.00, 5.00) \\   
 & Standard Deviation & (0.03, 0.05, 0.04, 0.04, 0.04, 0.04, 0.04, 0.05, 0.03, 0.05) \\
 & MSE($\boldsymbol{\Sigma}$) & 0.00 \\
\hline

\multicolumn{3}{c}{Simulation Setting 3 - UUU} \\
\hline
\multirow{4}{*}{\textbf{Group 1}}  & $\boldsymbol{\mu}$ & (4, 6, 4, 2, 2, 4, 6, 4, 6, 2) \\
 & $\boldsymbol{\hat{\mu}}$ & (4.00, 6.00, 4.00, 1.99, 2.00, 4.00, 5.98, 4.00, 6.00, 2.00) \\   
 & Standard Deviation & (0.05, 0.10, 0.06, 0.12, 0.07, 0.07, 0.09, 0.08, 0.09, 0.08) \\
 & MSE($\boldsymbol{\Sigma}$) & 0.01 \\
\hline
\multirow{4}{*}{\textbf{Group 2}}  & $\boldsymbol{\mu}$ & (5, 5, 3, 3, 7, 5, 3, 3, 7, 7) \\
 & $\boldsymbol{\hat{\mu}}$ & (5.00, 5.00, 2.99, 3.00, 7.00, 5.00, 3.00, 3.00, 7.00, 7.00) \\   
 & Standard Deviation & (0.04, 0.07, 0.07, 0.07, 0.04, 0.03, 0.08, 0.06, 0.06, 0.07) \\
 & MSE($\boldsymbol{\Sigma}$) & 0.02 \\
\hline
\multirow{4}{*}{\textbf{Group 3}}  & $\boldsymbol{\mu}$ & (2, 4, 4, 7, 2, 4, 7, 2, 7, 4) \\
 & $\boldsymbol{\hat{\mu}}$ & (1.99, 3.99, 3.99, 6.99, 1.98, 4.00, 7.01, 1.97, 6.99, 4.00) \\   
 & Standard Deviation & (0.09, 0.06, 0.08, 0.08, 0.09, 0.07, 0.09, 0.07, 0.10, 0.07) \\
 & MSE($\boldsymbol{\Sigma}$) & 0.01 \\
\hline

\end{tabular}
}
\end{table*}


For the simulation data analyses, the normalization factors representing library size estimate for samples were set to 1. All simulation analyses were performed with  {\sf R} version 4.2.2 Patched ``Innocent and Trusting", released on November 11, 2022, \citep{R2022}. 

We fitted all the proposed eight models from the MPLNFA family to all datasets for $G = 1 \text{ to } G_o+1$, where $G_o$ is the actual number of clusters and the number of latent factors ranged $K = 1 \text{ to } K_o+1$, where $K_o$ is the actual number of latent factors. All models were initialized with $k$-means initialization with three random starts.

The simulations demonstrated that our proposed method can recover the correct parameters for MPLNFA (see Table~\ref{tab-parameter-recovery}). The clustering results obtained using the BIC for model selection and the corresponding average ARI values are summarized in Table~\ref{table_g2modelselection}. As seen in Table~\ref{table_g2modelselection}, BIC selected the correct number of clusters, latent factors, and component scale matrix for the first simulation setting, with $M =$ UCC, with an average ARI value of 0.99. BIC provides similar results for settings 2 and 3 ($M =$ CCC and UUU, respectively), with an average ARI value of 1.00. 

\begin{table*}[h!]
\centering
\caption{Model selection results of the clusters (average ARI, standard deviation), latent factors and component scale matrices for each simulation setting compared to true values.}
\label{table_g2modelselection}
\resizebox{\textwidth}{!}{%
\begin{tabular}{@{\extracolsep{-9pt}}ccccccccccccccccccccccccc@{}}
\hline
 && \multicolumn{4}{c}{Cluster, $G$} & \hspace{5mm }& \hspace{5mm } & \multicolumn{4}{c}{Latent factor, $K$} &\hspace{5mm } & \hspace{5mm }& \multicolumn{4}{c}{Component scale matrix,  $M$} \\
\cline{3-6} \cline{9-12} \cline{15-18} 
Setting   && Original        & MPLNFA & mPoisson & mNB & \hspace{5mm }& \hspace{5mm }& Original & MPLNFA & mPoisson & mNB &\hspace{1mm } & & Original & MPLNFA & mPoisson & mNB \\ \hline
1 && 4 & 4  & 5 & 2 &\hspace{5mm }& \hspace{5mm }& 2     & 2   & NA & NA &\hspace{0.5 mm }& & UCC & UCC & NA & NA \\
  &&   & (0.99, 0.01) & (0.51, 0.08) &  (0.36, 0.19)  & \hspace{5mm }& \hspace{5mm }&    & &    &       &\hspace{0.5 mm } & &   &    &    & &    &  \\ 
2 && 2 & 2 & 3 & 2 & \hspace{5mm }& \hspace{5mm }& 3     & 3   & NA & NA & \hspace{0.5 mm }& & CCC & CCC  & NA & NA  \\
  &&   & (1.00, 0.00) & (1.00, 0.00) & (1.00, 0.00) & \hspace{5mm }& \hspace{5mm }&    & &   &      & \hspace{0.5 mm }& & &   &   & &    &  \\ 
3 && 3  & 3 & 4 & 3 & \hspace{5mm }& \hspace{5mm }& 4     & 4  & NA & NA &\hspace{0.5 mm }& & UUU & UUU  & NA & NA \\
  &&  & (1.00, 0.00) & (0.69, 0.02) & (0.52, 0.11) & \hspace{5mm }& \hspace{5mm }&      & & &      &\hspace{0.5 mm }& &  & & &  \\
 
\hline
\end{tabular}}
\end{table*}

Additionally, in Table~\ref{table_g2modelselection}, we clustered the same data with two other common approaches, also using BIC for model selection. The first, labelled `mPoisson' in Table~\ref{table_g2modelselection}, is a mixture of Poisson distributions via the ${\tt HTSCluster}$  {\sf R} package (version 2.0.10) \citep{rau2015}. This approach performed well in the fully constrained setting (i.e., simulation setting 2), with an average ARI of 1.00. However, it faltered with settings 1 and 3, where unconstrained elements were introduced. For settings 1 and 3, the average ARI of this approach was 0.51 and 0.69, respectively. Further, it consistently selected models with more groups than the actual number in the simulated datasets. These results were anticipated as a known limitation of this approach is that it cannot account for overdispersion, so to capture the larger variability, it will use additional components. The second approach is a mixture of Negative Binomial distributions, implemented via the ${\tt MBCluster.Seq}$  {\sf R} package (version 1.0) and labelled `mNB' in Table~\ref{table_g2modelselection} \citep{si2013}. This approach works very well on simulation setting 2, with an average ARI of 1.00 and identifies the correct number of groups, but it does not perform well on the other two settings. In Simulation setting 1, it has an average ARI of 0.36 and two fewer groups than the true number of groups, and in setting 3, it has an average ARI of 0.52 despite selecting the correct number of groups. A limitation of this approach is that it assumes independence between variables, and therefore cannot account for correlation, which greatly hinders its performance. As seen in these results, our proposed MPLNFA model performed the best for all simulation settings due to its ability to account for overdispersion and both positive and negative correlation, both of which are known limitations of the other two compared approaches.

\subsection{Transcriptome Data Analysis} 

To illustrate the applicability of the MPLNFA model, it was applied to a publicly available, curated RNA sequencing dataset of breast invasive carcinoma (\text{BRCA\_RNASeqGene-20160128}) from The Cancer Genome Atlas (TCGA) Program \citep{TCGAdataset2020}. Breast cancer is a commonly diagnosed cancer and a leading cause of cancer-related deaths in women worldwide \citep{WilkinsonGathani2022}. Clustering has been applied to understand breast cancer, by clustering of genes to identify co-expression networks, or clustering of tissues to identify subtypes \citep{sorlie2001gene, kreike2007gene}. The original dataset explored here had 878 patient tissues and 20,502 genes. The status of the estrogen receptor (ER), progesterone receptor (PR), and human epidermal growth factor receptor 2 (HER2) play a crucial role in breast cancer and its clinical management \citep{Chen2022}. Here, we focused only on a subset of tissues for which hormone receptor status was available and were females, resulting in a sample size of 712 tissues. Typically, only a subset of genes from the experiment are used for analysis purposes in order to reduce noise. The original number of 20,502 genes in the dataset was filtered to remove genes with missing expression values, resulting in 17,586 genes. The top 50 genes with the most variable gene expression across the 712 tissues were selected. The final dataset had a size of 712 tissues with 50 genes. 

In terms of patient characterization, cohort ages ranged from 26 to 90, with no age data available for 9\%. In terms of the pathologic stage, tissues from 11 stages (stages i, ia, ib, ii, iia, iib, iiia, iiib, iiic, iv, x) were present, with no data for 0.98\%. Staging quantifies the amount of cancer in the body, with a higher stage number indicating more cancer spread \citep{cuthrell2023breast}. Stage x indicates that the status cannot be determined. A majority of the cohort was identified as white (74\%), followed by black (8\%), Asian (7\%), American Indian or Alaska Native (0.15\%), with no data available for 10.85\%. In the cohort, 76.5\% were ER-positive and 66.7\% were PR-positive. In terms of HER2, 80.7\% were negative, 13.8\% were positive, 1.1\% were equivocal and 1.8\% did not have data. All patients were female. 

\begin{figure}[h!]
\centerline{\includegraphics[width=0.99\textwidth]{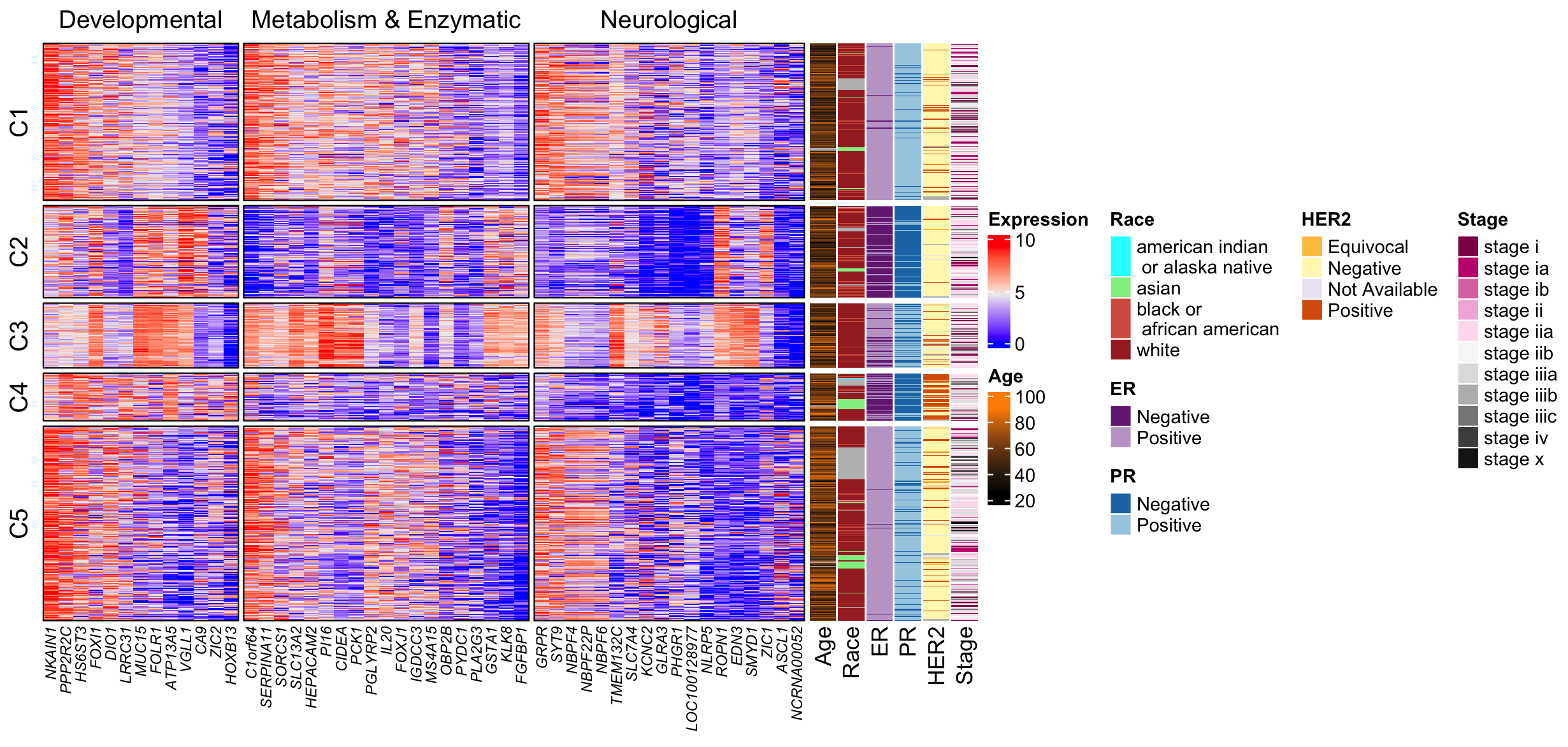}}
\caption{Heatmap showing, for Cluster 1 (C1) through Cluster 5 (C5), log-transformed gene expression patterns. This is for the $G = 5, k = 6$ and $M = \text{CUU}$ model selected by both BIC and ICL for the breast invasive carcinoma RNA-seq dataset (n = 712). The red and blue colours represent the expression levels, where red represents high expression and blue represents low expression. The rows represent the tissues and the columns represent the top $50$ most variable genes included in the study. ER is estrogen receptor, PR is progesterone receptor, and HER2 is human epidermal growth factor receptor 2.} \label{heatmapPlot}
\end{figure}

\begin{figure}[h!]
\centerline{\includegraphics[width=\textwidth]{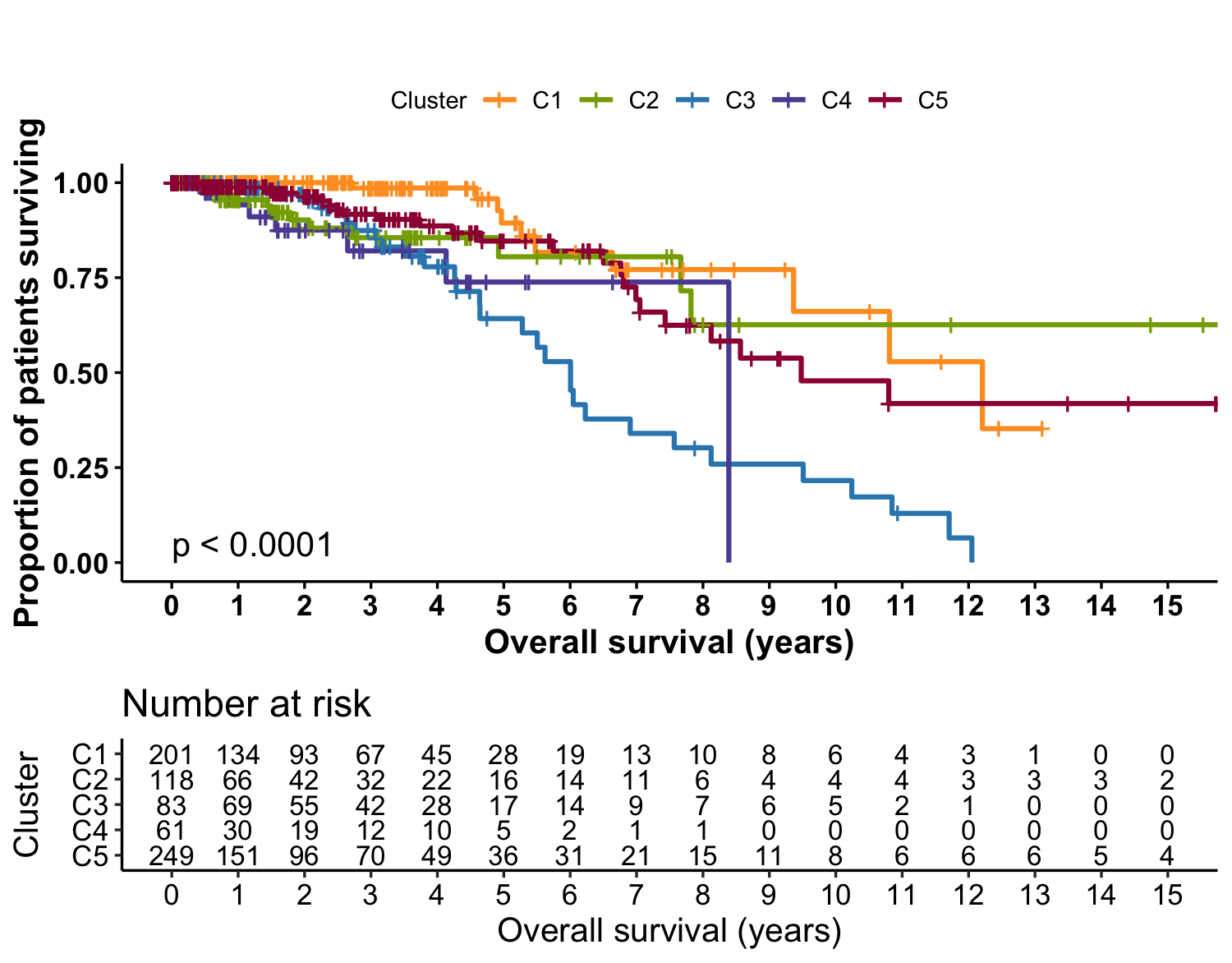}}
\caption{Kaplan-Meier plot of overall survival for Cluster 1 (C1) through Cluster 5 (C5) for the $G = 5, k = 6$ and $M = \text{CUU}$ model selected by both BIC and ICL for the breast invasive carcinoma RNA-seq dataset.} \label{survivalPlot}
\end{figure}

In order to identify subtypes of breast invasive carcinoma, the tissues were clustered for a clustering range from $G = 1 \text{ to } 8$, and a latent factor range of $k = 1 \text{ to } 10$, for all eight models listed in Table~\ref{tabfam}. All data analyses were performed using $k$-means initialization with $3$ runs. Both BIC and ICL selected a model with $G = 5, k = 6$ and $M = \text{CUU}$. The 5 clusters analyzed here are interchangeably referred to as the 5 subtypes of breast invasive carcinoma.  It was observed that the tissues were assigned to clusters with high posterior probabilities via the MPLNFA model (see Additional Figure 1). In this model, Clusters 1–5 were composed of $201 (28\%), 118 (17\%), 83 (12\%), 61 (8\%), \text{and } 249 (35\%)$ tissues, respectively. See Additional File 1 for the tissue composition of each cluster. In simulation studies, where the true labels are available, the ARI can be used to assess clustering performance. Because no true labels are available in real data, the predicted clusters were assessed in terms of overall survival, clinical characteristics and gene expression patterns. The distinct expression patterns between the 5 clusters are provided in Figure~\ref{heatmapPlot} \citep{heatmap2022}. The genetic subtypes were associated with a divergent overall survival (OS) in the overall cohort (P \textless 0.0001) as shown in Figure~\ref{survivalPlot}. The most favourable OS was observed for Cluster~2 and the most unfavourable OS was observed in Clusters~4, followed by ~3. In terms of race, age and pathologic stage, no significant associations were identified with the clusters. However, it was interesting to note that all races were represented in Cluster~4 and Cluster~4 also had the least proportion of tissues from stage i. 

The ER and PR positive status has shown to be associated with favourable patient outcomes \citep{davey2021impact}. Interestingly, Cluster~4 and Cluster~2, both contained the highest number of ER and PR negative cases. However, among all clusters, Cluster~4 contained the highest number of HER2 positive cases, while Cluster~2 had the most HER2 negative cases. HER2 overexpression or HER2 positive status in breast cancer has shown to be associated with poor prognosis and decreased survival \citep{Chen2022}. The PR is an estrogen-regulated gene, so ER-positive tumours are generally PR positive, and ER-negative tumours are PR negative \citep{bae2015poor}. This is observed across clusters in the current analysis. Further, the ER positive, PR positive status is known to have a better prognosis compared to ER negative, PR negative status \citep{bae2015poor}. This is evident for Cluster~1 and 5, where the majority of cases have both receptors positive.

To explore expression patterns, 50 genes in the dataset were divided into non-exclusive, three groups: developmental-related, metabolism and enzymatic-related or neurological-related genes. Cluster~1-3, and 5 showed variable expression levels among the three gene groups. However, Cluster~4 stood out as having high expression in developmental-related genes, followed by a mixed expression in metabolism-related genes, and low expression in neurological-related genes. The expression patterns observed here were supported in the literature. We selected some genes to illustrate this. For example, within development-related genes, \textit{NKAIN1} has high expression across all clusters. This gene encodes the sodium/potassium transporting ATPase interacting with 1 protein in humans. An association between high expression of \textit{NKAIN1} and breast cancer has been shown and \textit{NKAIN1} mRNA levels have been identified as a potential biomarker in the diagnosis of breast cancer \citep{AliHassanzadeh2022}. Further, \textit{HS6ST3} is the Heparan Sulfate 6-O-Sulfotransferase 3 protein-coding gene. It has been shown to be highly expressed in breast cancer cell lines and its silencing has diminished the tumour growth and progression \citep{Iravani2017}. Interestingly, Cluster~4 showed a high expression of \textit{HS6ST3}, Cluster~3 showed an intermediate level and Cluster~2 showed the lowest expression, in accordance with OS differences observed. 

In terms of metabolism-related genes, \textit{GSTA1} is the Glutathione S-Transferase Alpha 1 protein-coding gene. Reduced \textit{GSTA1} expression levels have been shown to be associated with increased breast cancer \citep{Jiyoung2006}, while higher expression has shown longer OS and disease-free survival \citep{Liu2020}. Here, Cluster~2 showed low expression of \textit{GSTA1} and Cluster~4 showed higher expression, agreeing with OS observed in the current study. The \textit{C1orf64} gene encodes the steroid receptor-associated and regulated protein. It has been shown to have a  high expression in breast tumours \citep{Naderi2017}. Here, Cluster~4 showed high expression of \textit{C1orf64} and Cluster~2 showed lower expression. In terms of neurological-related genes, \textit{ZIC1} is the zinc finger of the cerebellum 1 gene. It is known to have a tumour-suppressive role in breast cancer \citep{Han2018}. An increase in \textit{ZIC1} expression has been shown to suppress the growth of breast cancer cells and xenograft tumours \citep{Han2018}. Interestingly, Cluster~4 showed low expression of \textit{ZIC1} and Cluster~2 showed a high expression, supporting the OS differences observed. 

\section{Discussion}\label{sec:discussion}
A mixture of factor analyzers model for MPLN distribution as well as a family of mixture models based theorem is introduced. This is the first use of a mixture of MPLN factor analyzer distributions within the literature. To our knowledge, this is also the first use of a mixture of discrete factor analyzers within the literature. The proposed models are well-suited to high-dimensional applications as the number of scale parameters is linear in data dimensionality for all eight models, as opposed to in traditional MPLN, where the parameters grow quadratically. Further, a single MPLN factor analysis model can be obtained as a special case of the mixture of MPLNFA, i.e., with $G = 1$. Extensions applicable to factor analyzers model, such as the mixture of common factor analyzers \citep{Baek2010} can be applied to the MPLNFA family in future. A mixture of common factor analyzers is a restrictive form of the mixture of factor analyzers model and can be useful when the number of clusters and dimensionality are very large. 

Overall, applying MPLNFA identified 5 subtypes of breast invasive carcinoma that differ from each other in terms of patient survival outcomes, clinical covariates and gene expression. Connecting gene expression patterns with clinical covariates and patient outcomes is important in understanding disease diversity. Findings from our study provide a deeper understanding of the molecular and clinical heterogeneity of breast cancer and these may inform the development of subtype-specific treatment strategies to improve patient outcomes in future. 

\subsection*{Acknowledgments}
{\small This research was supported by the Postdoctoral Fellowship award from the Canadian Institutes of Health Research (Silva), the National Sciences and Engineering Research Council of Canada grant (2021-03812 - Subedi, 2023-06030 - McNicholas), and the Canada Research Chair Program (22020-00303 - Subedi, 2022-00494 - McNicholas).}

\subsection{Software availability}

The GitHub R package for this work is available at: \\
\texttt{https://github.com/anjalisilva/mixMPLNFA} and is released under the open source MIT license.

\appendix
\section{Appendix}
\subsection{Variational Approximations} \label{app:vae}

\noindent The marginal log-density of $\bY_i\mid Z_{ig}=1$ in Stage 1 can be written as:
\begin{align*}
\log f(\by_i\mid Z_{ig}=1) &= \int_{\real^d} \log f(\by_i\mid Z_{ig}=1) ~q(\bx_{ig}) ~d\bx_{ig}\\
&=  \int_{\real^d} \log \frac{f(\by_i,\bx_i\mid Z_{ig}=1)/q(\bx_{ig})}{f(\bx_i\mid\by_i,Z_{ig}=1)/q(\bx_{ig})} ~q(\bx_{ig}) ~d\bx_{ig}\\
&= \int_{\real^d} \left[\log f(\by_i,\bx_i\mid Z_{ig}=1) - \log q(\bx_{ig})\right] q(\bx_{ig}) ~ d\bx_{ig} + D_{KL}(q_{ig}\|f_{ig}) \\
&= F(q_{ig},\by_i) + D_{KL}(q_{ig}\|f_{ig}),
\end{align*}

\noindent The marginal log-density of $\bY_i\mid Z_{ig}=1$ in Stage 2 can be written as:
\begin{align*}
\log f(\by_i\mid Z_{ig}=1) &= \int_{\real^K} \int_{\real^d} \log f(\by_i\mid Z_{ig}=1) ~q_{\bx,\bu}(\bx_{ig},\bu_{ig}) ~d\bx_{ig}~d\bu_{ig}\\
&= \int_{\real^K} \int_{\real^d} \log \frac{f(\by_i,\bx_i,\bu_i\mid Z_{ig}=1)/q_{\bx,\bu}(\bx_{ig},\bu_{ig})}{f(\bx_i,\bu_i\mid\by_i,Z_{ig}=1)/q_{\bx,\bu}(\bx_{ig},\bu_{ig})} ~q_{\bx,\bu}(\bx_{ig},\bu_{ig}) ~d\bx_{ig}\\
&= \int_{\real^K} \int_{\real^d} \Big[\log f(\by_i,\bx_i,\bu_i\mid Z_{ig}=1) - \log q_{\bx,\bu}(\bx_{ig},\bu_{ig})\Big] \\
& \times q_{\bx,\bu}(\bx_{ig},\bu_{ig}) ~ d\bx_{ig} ~d\bu_{ig}+ D_{KL}(q_{ig}\|f_{ig}) \\
&= F(q_{ig},\by_i) + D_{KL}(q_{ig}\|f_{ig}),
\end{align*}

\noindent ELBO in Stage 2 under the assumption that $q_{x,u}(\bx,\bu)$ is factorable: 
\begin{align*}
F(q_{ig},\by_i)&=\int_{\real^K} \int_{\real^d} \Big[ \log f(\by_i \mid \bx_i,\bu_i, Z_{ig}=1) +\log f(\bx_i\mid \bu_i,Z_{ig}=1)+\log f(\bu_i\mid Z_{ig}=1)\\
&- \log q_{\bx}(\bx_{ig}) - \log q_{\bu}(\bu_{ig}) \Big] ~q_{\bx}(\bx_{ig})~q_{\bu}(\bu_{ig})~d\bx_{ig}~d\bu_{ig}\\
&=\int_{\real^K} \Bigg[\left\{\int_{\real^d} \Big[ \log f(\by_i \mid \bx_i,\bu_i, Z_{ig}=1) +\log f(\bx_i\mid \bu_i,Z_{ig}=1) \right.\\
& \left. -\log q_\bx(\bx_i)\Big]~q_{\bx}(\bx_{ig})~d\bx_{ig} \right\} + \log f(\bu_i\mid Z_{ig}=1) - \log q_{\bu}(\bu_{ig}) \Bigg] q_{\bu}(\bu_{ig}) ~d\bu_{ig}.
\end{align*}

\noindent Assuming that the approximating distributions (i.e. $q_\bx(\bx_{ig}) = N (\bmm_{ig},\mathbf{S}_{ig})$ and $q_\bu(\bu_{ig}) = N (\bP_{ig},\mathbf{Q}_{ig})$), the ELBO in Stage 2 becomes:
\begin{align*}
F(q_{ig},\by_i)&=\int_{\real^K} \Bigg[\bmm_{ig}'\by_i + \sum_{j=1}^d\left(\log k_j\right)y_{ij}-   \sum_{j=1}^d \left\{e^{\log k_j+m_{igj}+\frac{1}{2}S_{ig,jj}}+\log(y_{ij}!)\right\} \\
&+\frac{1}{2} \log |\bS_{ig}| +\frac{d}{2}-\frac{1}{2}(\bmm_{ig}-\bmu_g-\bLambda_g\bu_i)'\bPsi_g^{-1}(\bmm_{ig}-\bmu_g-\bLambda_g\bu_i)-\frac{1}{2}\tr\left(\bPsi_g^{-1}\mathbf{S}_{ig}\right)  \\
& -\frac{1}{2} \log |\bPsi_g| + \log f(\bu_i\mid Z_{ig}=1)  - \log q_{\bu}(\bu_{ig}) \Bigg] q_{\bu}(\bu_{ig}) ~d\bu_{ig}\\[3pt]
&=\bmm_{ig}'\by_i + \sum_{j=1}^d\left(\log k_j\right)y_{ij}-   \sum_{j=1}^d \left\{e^{\log k_j+m_{igj}+\frac{1}{2}S_{ig,jj}}+\log(y_{ij}!)\right\} +\frac{1}{2} \log |\bS_{ig}| +\frac{d}{2}\\
&-\frac{1}{2}(\bmm_{ig}-\bmu_{ig})^T\bPsi_g^{-1}(\bmm_{ig}-\bmu_{ig})+(\bmm_{ig}-\bmu_g)^T\bPsi_g^{-1}\bLambda_g\bP_{ig}-\frac{1}{2}\bP_{ig}^T\bLambda_g^T\bPsi_g^{-1}\bLambda_g\bP_{ig}\\
&-\frac{1}{2}\tr(\bLambda_g^T\bPsi_g^{-1}\bLambda_g\bQ_{ig})-\frac{1}{2}\tr\left(\bPsi_g^{-1}\mathbf{S}_{ig}\right)  -\frac{1}{2} \log |\bPsi_g|-\frac{1}{2}\bP_{ig}^T\bP_{ig}+\frac{1}{2}\log | \bQ_{ig}^{-1}|\\
&-\frac{1}{2}\tr( \bQ_{ig}) + \frac{K}{2}.
\end{align*}

\subsection{Family of Models}
\subsubsection{Isotropic assumption: $\bPsi_g = \psi_g \bI_d$}
\begin{align*}
l(\boldsymbol{\vartheta})&=\sum_{g=1}^G\sum_{i=1}^n z_{ig}\log \pi_g+ \sum_{g=1}^G\sum_{i=1}^n z_{ig}\log f(\by_{i}\mid Z_{ig}=1, \bmu_g,\bPsi_g,\bLambda_g).\\
&=\sum_{g=1}^G\sum_{i=1}^n z_{ig}\log \pi_g+ \sum_{g=1}^G\sum_{i=1}^n z_{ig} \left \{F(q_{ig},\by_i) + D_{KL}(q_{ig}\mid\mid f_{ig})\right \}
\end{align*}
Under the isotropic constraint (i.e. $\bPsi_g = \psi_g \bI_d$),
\begin{align*}
\sum_{g=1}^G\sum_{i=1}^nz_{ig}F(q_{ig},\by_i)&=\sum_{g=1}^G\sum_{i=1}^nz_{ig}\left[-\frac{\psi_g^{-1}}{2}(\bmm_{ig}-\bmu_{ig})^T(\bmm_{ig}-\bmu_{ig})+\psi_g^{-1}(\bmm_{ig}-\bmu_g)^T\bLambda_g\bP_{ig}  \right.\\
&\left.+\frac{d}{2} \log \psi_g^{-1}-\frac{\psi_g^{-1}}{2}\bP_{ig}^T\bLambda_g^T\bLambda_g\bP_{ig}-\frac{\psi_g^{-1}}{2}\tr(\bLambda_g^T\bLambda_g\bQ_{g})-\frac{\psi_g^{-1}}{2}\tr\left(\mathbf{S}_{ig}\right)  \right]+C,\\[3pt]
&=\sum_{g=1}^G\left[-\frac{\psi_g^{-1}}{2}n_g\tr(\bW_{g})+\psi_g^{-1}n_g\tr(\bW_g\bLambda_g\bbeta_g)+\frac{dn_g}{2}\log \psi_g^{-1}\right.\\
&\left.-\frac{\psi_g{^{-1}}n_g}{2}\tr(\bLambda_g^T\bLambda_g\bQ_{g}) -\frac{\psi_g{^{-1}}}{2}n_g\tr(\bbeta^T\bLambda_g^T\bLambda_g\bbeta_g\bW_g) \right.\\
& \left. - \frac{\psi_g^{-1}}{2}\tr\left(\sum_{i=1}^nz_{ig}\mathbf{S}_{ig}\right)\right]+C,
\end{align*}
where $C$ is a constant with respect to $\psi_g^{-1}$ and $n_g=\sum_{i=1}^nz_{ig}$. Taking derivative with respect to $\psi^{-1}_g$ and setting it to 0, we get
\begin{align*}
d\hat{\psi}_g&-\tr(\bW_g)+2\tr(\bW_g\hat{\bLambda}_g\hat{\bbeta}_g)-\tr(\hat{\bbeta}^T\hat{\bLambda}_g^T\hat{\bLambda}_g\hat{\bbeta}_g\bW_g)\\
&-\tr(\hat{\bLambda}_g^T\hat{\bLambda}_g\bQ_{g})-
\tr\left(\frac{\sum_{i=1}^nz_{ig}\mathbf{S}_{ig}}{n_g}\right)=0.
\end{align*}
Simplifying this, we get
\begin{equation}
\hat{\psi}_g=\frac{1}{d}\tr\left(\bW_g-\hat{\bLambda}_g\hat{\bbeta}_g\bW_g+\frac{\sum_{i=1}^nz_{ig}\mathbf{S}_{ig}}{n_g}\right).
\end{equation}

\subsubsection{Equal Variance: $\bPsi_g = \bPsi$}
Under the equal variance constraint (i.e. $\bPsi_g = \bPsi$),
\begin{align*}
\sum_{g=1}^G\sum_{i=1}^nz_{ig}F(q_{ig},\by_i)&=\sum_{g=1}^G\sum_{i=1}^nz_{ig}\left[-\frac{1}{2}(\bmm_{ig}-\bmu_{ig})^T\bPsi^{-1}(\bmm_{ig}-\bmu_{ig})\right.\\
&\left.+(\bmm_{ig}-\bmu_g)^T\bPsi^{-1}\bLambda_g\bP_{ig}+\frac{1}{2} \log |\bPsi^{-1}|-\frac{1}{2}\bP_{ig}^T\bLambda_g^T\bPsi^{-1}\bLambda_g\bP_{ig}\right. \\
&\left. -\frac{1}{2}\tr(\bLambda_g^T\bPsi^{-1}\bLambda_g\bQ_{g})-\frac{1}{2}\tr\left(\bPsi^{-1}\mathbf{S}_{ig}\right)  \right]+C,\\[3pt]
&=\sum_{g=1}^G\left[-\frac{n_g}{2}\tr(\bPsi^{-1}\bW_{g})+n_g\tr(\bW_g\bPsi^{-1}\bLambda_g\bbeta_g)-\frac{n_g}{2}\tr(\bLambda_g^T\bPsi^{-1}\bLambda_g\bQ_{g})\right.\\
&\left.+\frac{n_g}{2}\log |\bPsi^{-1}|-\frac{\psi_g{^{-1}}}{2}n_g\tr(\bbeta^T\bLambda_g^T\bPsi^{-1}\bLambda_g\bbeta_g\bW_g)\right.\\
&\left.-\frac{1}{2}\tr\left(\bPsi^{-1}\sum_{i=1}^nz_{ig}\mathbf{S}_{ig}\right)\right]+C,
\end{align*}
where $C$ is a constant with respect to $\bPsi$ and $n_g=\sum_{i=1}^nz_{ig}$. Taking derivative with respect to $\bPsi^{-1}$ and setting it to 0, we get
\begin{align*}
n\hat{\bPsi}&-\sum_{g=1}^Gn_g\bW_g+2\sum_{g=1}^Gn_g\hat{\bLambda}_g\hat{\bbeta}_g\bW_g-\sum_{g=1}^Gn_g\hat{\bbeta}^T\hat{\bLambda}_g^T\hat{\bLambda}_g\hat{\bbeta}_g\bW_g\\
&-\sum_{g=1}^Gn_g\hat{\bLambda}_g^T\hat{\bLambda}_g\bQ_{g}-\tr\left(\frac{\sum_{i=1}^nz_{ig}\mathbf{S}_{ig}}{n_g}\right)=0.
\end{align*}
Simplifying this, we get
\begin{align*}
\hat{\bPsi}&=\diag\left(\sum_{g=1}^G\frac{n_g}{n}\bW_g-\sum_{g=1}^G\frac{n_g}{n}\hat{\bLambda}_g\hat{\bbeta}_g\bW_g+\frac{1}{n}\sum_{g=1}^G\sum_{i=1}^nz_{ig}\mathbf{S}_{ig}\right).\\
&=\diag\left(\sum_{g=1}^G\frac{n_g}{n}\bW_g-\sum_{g=1}^G\frac{n_g}{n}\hat{\bLambda}_g\hat{\bbeta}_g\bW_g+\frac{1}{n}\sum_{g=1}^G\sum_{i=1}^nz_{ig}\mathbf{S}_{ig}\right)
\end{align*}

\subsubsection{Equal Variance and isotropic constraint $\bPsi_g=\psi\bI_d$}
\begin{align*}
\sum_{g=1}^G\sum_{i=1}^nz_{ig}F(q_{ig},\by_i)&=\sum_{g=1}^G\sum_{i=1}^nz_{ig}\left[-\frac{\psi^{-1}}{2}(\bmm_{ig}-\bmu_{ig})^T(\bmm_{ig}-\bmu_{ig})+\psi^{-1}(\bmm_{ig}-\bmu_g)^T\bLambda_g\bP_{ig}  \right.\\
&\left.+\frac{d}{2} \log \psi^{-1}-\frac{\psi^{-1}}{2}\bP_{ig}^T\bLambda_g^T\bLambda_g\bP_{ig}-\frac{\psi^{-1}}{2}\tr(\bLambda_g^T\bLambda_g\bQ_{g})-\frac{\psi^{-1}}{2}\tr\left(\mathbf{S}_{ig}\right)  \right]+C,\\[3pt]
&=\sum_{g=1}^G\left[-\frac{\psi^{-1}}{2}n_g\tr(\bW_{g})+\psi^{-1}n_g\tr(\bW_g\bLambda_g\bbeta_g)+\frac{dn_g}{2}\log \psi^{-1}\right.\\
&\left.-\frac{\psi{^{-1}}n_g}{2}\tr(\bLambda_g^T\bLambda_g\bQ_{g})-\frac{\psi{^{-1}}}{2}n_g\tr(\bbeta^T\bLambda_g^T\bLambda_g\bbeta_g\bW_g)\right.\\
&\left.-\frac{\psi^{-1}}{2}\tr\left(\sum_{i=1}^nz_{ig}\mathbf{S}_{ig}\right)\right]+C,
\end{align*}
where $C$ is a constant with respect to $\psi^{-1}$ and $n_g=\sum_{i=1}^nz_{ig}$. Taking derivative with respect to $\psi^{-1}_g$ and setting it to 0, we get
 \begin{align*}
dn\hat{\psi}&-\tr(\sum_{g=1}^Gn_g\bW_g)+ 2\tr(\sum_{g=1}^Gn_g\bW_g\hat{\bLambda}_g\hat{\bbeta}_g)-\tr(\sum_{g=1}^Gn_g\hat{\bbeta}^T\hat{\bLambda}_g^T\hat{\bLambda}_g\hat{\bbeta}_g\bW_g)\\
&-\tr(\sum_{g=1}^Gn_g\hat{\bLambda}_g^T\hat{\bLambda}_g\bQ_{g})-\tr\left(\sum_{g=1}^G\sum_{i=1}^nz_{ig}\mathbf{S}_{ig}\right)=0.
\end{align*}

And, simplifying this, we get
\begin{align}
\hat{\psi}&=\frac{1}{d}\tr\left(\sum_{g=1}^G\frac{n_g}{n}\bW_g-\sum_{g=1}^G\frac{n_g}{n}\hat{\bLambda}_g\hat{\bbeta}_g\bW_g+\frac{1}{n}\sum_{g=1}^G\sum_{i=1}^nz_{ig}\mathbf{S}_{ig}\right)
\end{align}

\subsubsection{Equal Loading Matrices: $\bLambda_g = \bLambda$}
Under the equal loading matrices constraint (i.e. $\bLambda_g = \bLambda$),
\begin{align*}
\sum_{g=1}^G\sum_{i=1}^nz_{ig}F(q_{ig},\by_i)&=
\sum_{g=1}^G\sum_{i=1}^nz_{ig}\left((\bmm_{ig}-\bmu_g)^T\bPsi_g^{-1}\bLambda\bP_{ig}-\frac{1}{2}\bP_{ig}^T\bLambda^T\bPsi_g^{-1}\bLambda\bP_{ig}\right.\\
&\left.-\frac{1}{2}\tr(\bLambda^T\bPsi_g^{-1}\bLambda\bQ_{g})\right)\\
&=\sum_{g=1}^G\left(n_g\tr\left(\bbeta_g\bW_g\bPsi_g^{-1}\bLambda\right)-\frac{n_g}{2}\tr\left(\bbeta_g^T\bLambda^T\bPsi_g^{-1}\bLambda\bbeta_g\bW_{g}\right)\right.\\
&\left. -\frac{n_g}{2}tr(\bLambda^T\bPsi_g^{-1}\bLambda\bQ_{g})\right).
\end{align*}
Taking derivative with respect to $\bLambda$ and setting it equal to 0, we get
\begin{align*}
0&=\sum_{g=1}^Gn_g \hat{\bPsi}_g^{-1}\bW_g\hat{\bbeta}_g^T-\sum_{g=1}^Gn_g\hat{\bPsi}_g^{-1}\hat{\bLambda}\hat{\bbeta}_g\bW_g\hat{\bbeta}_g^T-\sum_{g=1}^G n_g\hat{\bPsi}_g^{-1}\hat{\bLambda}\bQ_{g}\\
&=\sum_{g=1}^G n_g\hat{\bPsi}_g^{-1}\bW_g\hat{\bbeta}_g^T-\sum_{g=1}^Gn_g\hat{\bPsi}_g^{-1}\hat{\bLambda}\left(\hat{\bbeta}_g\bW_g\hat{\bbeta}_g^T-\bI_p-\hat{\bbeta}_g\hat{\bLambda}\right)\\
&=\sum_{g=1}^Gn_g \hat{\bPsi}_g^{-1}\bW_g\hat{\bbeta}_g^T-\sum_{g=1}^Gn_g\hat{\bPsi}_g^{-1}\hat{\bLambda}\hat{\bTheta}_g.
\end{align*}
In this case, the loading matrix cannot be solved directly and must be solved in a row-by-row manner as suggested by \cite{mcnicholas2008}. Hence,
\begin{align}
\hat{\lambda}_i&=\mathbf{r}_i\left(\sum_{g=1}^G\frac{n_g}{\hat{\Psi}_{g,(i)}} \hat{\bTheta}_g\right)^{-1},
\end{align}
where $\lambda_i$ is the $i^{th}$ row of the matrix $\bLambda$, $\Psi_{g,(i)}$ is the $i^{th}$ diagonal element of $\bPsi_g$ and $\mathbf{r}_i$ is the $i^{th}$ row of the matrix $\sum_{g=1}^Gn_g\hat{\bPsi}_g^{-1}\bW_g\hat{\bbeta}_g^T$.

\newpage
\subsection{Parameter Updates} \label{app:para-update}
\begin{longtable}{p{.07\textwidth}  p{.15\textwidth}p{.70\textwidth}} 
\caption{Parameter Updates for the family of models based on constraints on $\bLambda_g$ and $\bPsi_g$.}
\label{tab_para-updates} \\
\hline
Model & Parameters & Estimates \\
\hline
\endfirsthead

\hline
Model & Parameters & Estimates \\
\hline 
\endhead

 &&Continued on next page...\\ \hline
\endfoot

\endlastfoot

UUU &$\bbeta_g$& $\hat{\bbeta}_g=\hat{\bLambda}_g^T(\hat{\bLambda}_g\hat{\bLambda}_g^T+\bPsi_g)^{-1}$\\
&$\bTheta_g$   & $\hat{\bTheta}_g=\bI_K -\hat{\bbeta_g}\hat{\bLambda}_g+\hat{\bbeta}_g\bW_g\hat{\bbeta_g}^T$\\
&$\bLambda_g$& $\hat{\bLambda}_g=\bW_g\hat{\bbeta}^T\widehat{\bTheta}_g^{-1}$\\
&$\bPsi_g$&$\hat{\bPsi}_g=\diag\left(\bW_g-\hat{\bLambda}_g\hat{\bbeta}_g\bW_g+\frac{\sum_{i=1}^n \widehat{Z}_{ig}\bS_{ig}}{\sum_{i=1}^n\widehat{Z}_{ig}}\right)$ \\
\\
UUC&$\bbeta_g$& $\hat{\bbeta}_g=\hat{\bLambda}_g^T(\hat{\bLambda}_g\hat{\bLambda}_g^T+\psi_g\bI_d)^{-1}$\\
&$\bTheta_g$   & $\hat{\bTheta}_g=\bI_K -\hat{\bbeta_g}\hat{\bLambda}_g+\hat{\bbeta}_g\bW_g\hat{\bbeta_g}^T$\\
&$\bLambda_g$& $\hat{\bLambda}_g=\bW_g\hat{\bbeta}^T\widehat{\bTheta}_g^{-1}$\\
&$\bPsi_g=\psi_g\bI_d$&$\hat{\psi}_g=\frac{1}{d}\tr\left(\bW_g-\hat{\bLambda}_g\hat{\bbeta}_g\bW_g+\frac{\sum_{i=1}^nz_{ig}\mathbf{S}_{ig}}{n_g}\right)$\\
\\
UCU & $\bbeta_g$& $\hat{\bbeta}_g=\hat{\bLambda}_g^T(\hat{\bLambda}_g\hat{\bLambda}_g^T+\bPsi)^{-1}$\\
&$\bTheta_g$   & $\hat{\bTheta}_g=\bI_K -\hat{\bbeta_g}\hat{\bLambda}_g+\hat{\bbeta}_g\bW_g\hat{\bbeta_g}^T$\\
&$\bLambda_g$& $\hat{\bLambda}_g=\bW_g\hat{\bbeta}^T\widehat{\bTheta}_g^{-1}$\\
&$\bPsi_g=\bPsi$&$\hat{\bPsi}=\diag\left(\sum_{g=1}^G\frac{n_g}{n}\bW_g-\sum_{g=1}^G\frac{n_g}{n}\hat{\bLambda}_g\hat{\bbeta}_g\bW_g+\frac{1}{n}\sum_{g=1}^G\sum_{i=1}^nz_{ig}\mathbf{S}_{ig}\right)$ \\ 
\\
UCC & $\bbeta_g$& $\hat{\bbeta}_g=\hat{\bLambda}_g^T(\hat{\bLambda}_g\hat{\bLambda}_g^T+\psi\bI)^{-1}$\\
&$\bTheta_g$   & $\hat{\bTheta}_g=\bI_K -\hat{\bbeta_g}\hat{\bLambda}_g+\hat{\bbeta}_g\bW_g\hat{\bbeta_g}^T$\\
&$\bLambda_g$& $\hat{\bLambda}_g=\bW_g\hat{\bbeta}^T\widehat{\bTheta}_g^{-1}$\\
&$\bPsi_g=\psi\bI$&$\hat{\psi}=\frac{1}{d}\tr\left(\sum_{g=1}^G\frac{n_g}{n}\bW_g-\sum_{g=1}^G\frac{n_g}{n}\hat{\bLambda}_g\hat{\bbeta}_g\bW_g+\frac{1}{n}\sum_{g=1}^G\sum_{i=1}^nz_{ig}\mathbf{S}_{ig}\right)$ \\
\\
CUU &$\bbeta_g$& $\hat{\bbeta}_g=\hat{\bLambda}^T(\hat{\bLambda}\hat{\bLambda}^T+\bPsi_g)^{-1}$\\
&$\bTheta_g$   & $\hat{\bTheta}_g=\bI_K -\hat{\bbeta_g}\hat{\bLambda}+\hat{\bbeta}_g\bW_g\hat{\bbeta_g}^T$\\
&$\bLambda_g=\bLambda$&  Updated using Equation \ref{common_lambda}.\\[3pt]
&$\bPsi_g$&$\hat{\bPsi}_g=\diag\left(\bW_g-\hat{\bLambda}\hat{\bbeta}_g\bW_g+\frac{\sum_{i=1}^n \widehat{Z}_{ig}\bS_{ig}}{\sum_{i=1}^n\widehat{Z}_{ig}}\right)$ \\
\\
CUC &  $\bbeta_g$& $\hat{\bbeta}_g=\hat{\bLambda}^T(\hat{\bLambda}\hat{\bLambda}^T+\psi_g\bI_d)^{-1}$\\
&$\bTheta_g$   & $\hat{\bTheta}_g=\bI_K -\hat{\bbeta_g}\hat{\bLambda}+\hat{\bbeta}_g\bW_g\hat{\bbeta_g}^T$\\
&$\bLambda_g=\bLambda$&  Updated using Equation \ref{common_lambda}.\\
&$\bPsi_g=\psi_g\bI_d$&$\hat{\psi}_g=\frac{1}{d}\tr\left(\bW_g-\hat{\bLambda}\hat{\bbeta}_g\bW_g+\frac{\sum_{i=1}^nz_{ig}\mathbf{S}_{ig}}{n_g}\right)$\\
\\
CCU &  $\bbeta_g=\bbeta$& $\hat{\bbeta}=\hat{\bLambda}^T(\hat{\bLambda}\hat{\bLambda}^T+\bPsi)^{-1}$\\
&$\bTheta_g$   & $\hat{\bTheta}_g=\bI_K -\hat{\bbeta}\hat{\bLambda}+\hat{\bbeta}\bW_g\hat{\bbeta}^T$\\
&$\bLambda_g=\bLambda$&  Updated using Equation \ref{common_lambda}.\\
&$\bPsi_g=\bPsi$&$\hat{\bPsi}=\diag\left(\sum_{g=1}^G\frac{n_g}{n}\bW_g-\sum_{g=1}^G\frac{n_g}{n}\hat{\bLambda}\hat{\bbeta}\bW_g+\frac{1}{n}\sum_{g=1}^G\sum_{i=1}^nz_{ig}\mathbf{S}_{ig}\right)$ \\ 
\\
CCC & $\bbeta_g=\bbeta$& $\hat{\bbeta}=\hat{\bLambda}^T(\hat{\bLambda}\hat{\bLambda}^T+\psi\bI_d)^{-1}$\\
&$\bTheta_g$   & $\hat{\bTheta}_g=\bI_K -\hat{\bbeta}\hat{\bLambda}+\hat{\bbeta}\bW_g\hat{\bbeta}^T$\\
&$\bLambda_g=\bLambda$&  Updated using Equation \ref{common_lambda}.    \\
&$\bPsi_g=\psi\bI$&$\hat{\psi}=\frac{1}{d}\tr\left(\sum_{g=1}^G\frac{n_g}{n}\bW_g-\sum_{g=1}^G\frac{n_g}{n}\hat{\bLambda}\hat{\bbeta}\bW_g+\frac{1}{n}\sum_{g=1}^G\sum_{i=1}^nz_{ig}\mathbf{S}_{ig}\right)$ \\
\hline
\end{longtable}

%

\end{document}